\documentclass[11pt,preprint]{aastex}
\usepackage{apjfonts}
\usepackage[pdfpagemode={UseOutlines},bookmarks,bookmarksopen,colorlinks,linkcolor={black},citecolor={black},urlcolor={black}]{hyperref}
\usepackage{graphicx}
\usepackage{color}

\begin{document}
\shorttitle{Torques on Misaligned Disks}
\title{Tidal Torques on Misaligned Disks in Binary Systems}
\shortauthors{\sc{Lubow, Martin \& Nixon}} 
\author{Stephen~H.~Lubow\altaffilmark{1}, Rebecca~G.~Martin\altaffilmark{2,3} \&  Chris~Nixon\altaffilmark{2,4}} 
\affil{
$^{1}$Space Telescope Science Institute, 3700 San Martin Drive, Baltimore, MD 21218, USA\\
$^{2}$JILA, University of Colorado \& NIST, Boulder CO 80309-0440, USA\\
$^{3}$Sagan Fellow\\
$^{4}$Einstein Fellow
}

\begin{abstract}
We extend previous studies of the tidal truncation of coplanar  disks in binary systems to the more general case of noncoplanar disks. 
As in the prograde coplanar case, Lindblad resonances play a key role in tidal truncation.
We analyze the tidal torque acting on a misaligned nearly circular disk in a circular orbit binary system. We concentrate on the 2:1 inner Lindblad resonance associated with the $m=2$ tidal forcing (for azimuthal wavenumber $m$) that plays a major role in the usual coplanar case. We determine the inclination dependence of this torque, which is approximately $\cos^8{(i/2)}$ for misalignment angle $i$. Compared to the prograde coplanar case  ($i=0$), this torque decreases by a factor of about 2 for $i = \pi/6$ and by a factor of about 20 for  $i=\pi/2$. The Lindblad torque decreases to zero for a tilt angle of $\pi$  (counter-rotation), consistent with previous investigations. The effects of higher order resonances associated with $m>2$ tidal forcing may contribute somewhat, but are much more limited than in the $i=0$ case. These results suggest that misaligned disks in binary systems can be significantly extended compared to their coplanar counterparts. In cases where a disk is sufficiently inclined and viscous, it can overrun all Lindblad resonances and overflow the Roche lobe of the disk central object.
\end{abstract}

\keywords{accretion, accretion disks --- binaries: general --- hydrodynamics --- black hole physics}

\section{Introduction}
\label{intro}
Accretion disks occur in many types of astrophysical binaries on all scales, from circumplanetary disks to accretion in supermassive black hole binaries. Over this wide range of scales, the disk size is thought to be limited by the tidal torques exerted by the companion.  For accretion to occur in a disk, viscous torques transport angular momentum outwards causing the disk to expand \citep{Pringle81}. To truncate the disk, the tidal torques must dominate over the viscous torques \citep{Goldreich80}. The tidal torques transfer angular momentum from the disk to the orbit of the binary. These torques provide the disk angular momentum loss that permits the accretion of gas onto the central object without further expansion beyond the tidal truncation radius. 

Resonances in the disk play a critical role in providing the tidal torque (Goldreich \& Tremaine 1979, 1980). The disk size is determined by the location of the innermost resonance in the disk whose torque is capable of overcoming the viscous torque. Previous studies of the tidal truncation of disks in binaries have considered both circular orbit and eccentric orbit binaries (e.g., \citealt{Paczynski77}, \citealt{Papaloizou77}, \citealt{Artymowicz94}). They assumed that the disk lies in the orbit plane of the binary and rotates in the same prograde sense as the binary. In this paper, we consider the tidal truncation of (nearly) circular disks that are misaligned with the orbital plane of circular orbit binaries.

For a circular orbit binary and a coplanar prograde disk, the dominant contribution to the torque is associated with the azimuthal wavenumber $m=2$ tidal  distortions \citep{Papaloizou77}. These distortions give rise to an inner Lindblad resonance where the disk orbital frequency is twice that of the binary (2:1 resonance). This resonance is the closest significant Lindblad resonance to central object about which the disk orbits (the $m=1$ resonance occurs very near the central object, but the tidal forcing there is weak). In the case of a binary with mass ratio of order unity, this resonance is so powerful that it can truncate the disk at a smaller radius than the resonance location. This occurs because the resonance has a characteristic radial width. The torque resulting from the tail of this resonance is sufficiently strong to cause disk truncation.

By ignoring pressure effects in the coplanar case, one can model the disk streamlines as simple periodic ballistic particle orbits in the frame of the binary. Due to the strong tidal distortions, these orbits can strongly cross at some distance from the central object for binaries of order unity mass ratios. The orbits crossings lead to strong tidal dissipation that then provide a limit to the disk size (\citealt{Paczynski77}, \citealt{Papaloizou77}). In a more accurate treatment of the disk as a fluid, the  effect of a resonance is to launch waves that carry angular momentum. As the waves damp, they exert a torque on the disk that can effect its size. Hydrodynamical simulations of disks in binaries exhibit two-armed spiral waves that are caused by the $m=2$ Lindblad resonance (e.g., \citealt{Savonije94}, \citealt{Martin11}).

Disks are often expected to be coplanar and prograde with respect to the binary orbit. For example, disks in mass exchange binaries are naturally coplanar and prograde due to the coplanarity and angular momentum of the gas stream that feeds it. Tidal torques are expected to bring disks into alignment (\citealt{Papaloizou95}; \citealt{Larwood96}; \citealt{Bate00}; \citealt{Lubow00}). But in many cases, disks are expected to be misaligned with respect to the binary plane due to initial conditions or instabilities (e.g. \citealt{Wijers99}; \citealt{Ogilvie01}; \citealt{Begelman06}; \citealt{Bate10}; \citealt{Nixon14}). Furthermore, there is observational evidence of strong disk misalignment in young binaries (e.g., \citealt{Jensen14}).

There has been little attention paid to the effect of a misalignment on the truncation torque. \cite{Larwood96} report that there is a (small) change in the tidal truncation radius for moderately misaligned disks based on SPH simulations. \cite{Martin14a} reported a noticeable increase in disk radius for strongly misaligned eccentric disks in SPH simulations. However, for the maximally misaligned case of a counter-rotating disk in a circular orbit binary, the usual Lindblad resonance condition predicts that no resonances lie within the disk. Therefore, the tidal torque arising from Lindblad resonances should be zero \citep{Nixon11a}. In such a case, the disk may expand until the potential is no longer able to support circular orbits, probably transferring mass back to the companion. In this paper, we explore how the tidal torque varies with misalignment angle.

We derive the inclination dependence of the tidal truncation torque (Section~\ref{torque}) and compare this to the viscous torque to determine when the disk may be significantly extended (Section~\ref{comp}).

\section{Lindblad torque from a misaligned perturber}
\label{torque}
 
\subsection{Lindblad resonances}
We explore the effect of a misaligned perturber of mass $M_{\rm p}$ on a disk that is in orbit about the other binary component of mass $M_{\rm c}$ that lies central to a disk. Consider a cylindrical coordinate system $(r, \theta, z)$, whose origin lies on the disk central object and where $z=0$ lies at the disk midplane.  The disk rotates with the increasing $\theta$ and so its orbital frequency $\Omega >0$. Following \cite{Goldreich79}, we decompose the potential as
\begin{equation}
\Phi(r, \theta, z, t) = \sum_{\ell, m} \Phi_{\ell, m}(r,z) \cos{(m \theta - \ell \Omega_{\rm b} t)}\,,
\label{Phi}
\end{equation}
where the binary orbital frequency is $\Omega_{\rm b}>0$, $m \ge 0$  and $\ell$ ranges over all integers (negative, zero, and positive). To determine the individual potential components $\Phi_{\ell, m}$ for $\ell \ne 0$ and $m>0$, we invert Equation (\ref{Phi}) to obtain
\begin{equation}
\Phi_{\ell,m} = \frac{1}{2{\rm \pi}^2} \int^{2{\rm \pi}}_0 d(\Omega_{\rm b} t) \int_0^{2{\rm \pi}}  d\theta \, \Phi(r, \theta, z, t)\cos(m\theta - \ell\Omega_{\rm b} t)\,.
\label{Philm}
\end{equation}
It can be shown that for a circular orbit binary (as we assume in this paper), $\Phi_{\ell,m}$ is nonzero only for $|\ell| = m$. If the binary were to reside on a coplanar prograde (retrograde) orbit with respect to the disk, the nonzero components of the potential would all have positive (negative) $\ell$ values.

In this paper, we consider the torques due to Lindblad resonances in a disk. For a Keplerian disk, they occur at radii
$r$ that satisfy
\begin{equation}
\Omega(r) = \frac{\ell \Omega_{\rm b}}{m \mp 1},
\label{res}
\end{equation}
where the upper (lower) sign is for an inner (outer) Lindblad resonance. For a circular orbit binary with $m >1$, no Lindblad torques occur for $\ell <0$ because the Lindblad resonance condition (\ref{res})  can only be satisfied for positive $\ell$ values, since $\Omega(r) >0$.

We consider the  $\ell  = m$ potential component that is responsible for the resonant torque. The binary orbital separation is $a$, the binary orbit plane is inclined with respect to the disk plane by angle $i$, the angle of the perturber orbit (argument of periapsis) is  $\phi = \Omega_{\rm b} t$, and $\theta=0$ is along the line of ascending node of the perturber. At the disk midplane, $z=0$, the tidal potential in the disk due to the perturber is given by
\begin{equation}
\Phi(x, i, \theta, \phi)=\frac{-G M_{\rm p}/a}{\sqrt{x^2 - 2 x [\cos{(\phi)} \cos{(\theta)} + \cos{(i)} \sin{(\phi)}\sin{(\theta)}]+1}}\,,
\label{pi}
\end{equation}
where $x=r/a$. We then evaluate $\Phi_{m,m}$ at the disk midplane as
\begin{equation}
\Phi_{m,m}(x, i) = \frac{1}{2{\rm \pi}^2} \int^{2{\rm \pi}}_0 d \phi \int_0^{2{\rm \pi}}  d\theta \, \, \Phi(x, i, \theta, \phi) \cos(m(\theta - \phi)).
\label{Phimme}
\end{equation}

Using the potential component $\Phi_{m,m}(x,i)$, we determine the resonant torque on the disk $T_{m,m}$ at the inner Lindblad resonance based on Equation (46) of \cite{Goldreich79} 
\begin{equation}
\label{gt}
T_{m,m}(i) = -{\rm \pi}^2\Sigma(r) \frac{\left( 2 m \Phi_{m,m} + x \partial_x \Phi_{m,m} \right)^2}{ 3 \Omega \Omega_{\rm b}}\,,
\label{tgt}
\end{equation}
for disk surface density $\Sigma$, where all radially dependent quantities are evaluated at the resonance radius $r=r_{\rm res}=(1-1/m)^{2/3} (1-\mu)^{1/3} a$ and $\mu = M_{\rm p}/(M_{\rm p}+M_{\rm c})$ is the mass fraction of the perturber. In deriving this equation, we assumed that the unperturbed orbital motion at the resonance radius is Keplerian about the object of mass $M_{\rm c}$.

\subsection{\texorpdfstring{$m=2$}{m=2} Lindblad resonance}
We consider the  $\ell  = m = 2$ potential component that is responsible for the resonant torque at the 2:1 inner Lindblad resonance. For $x <1$, we apply Equation (\ref{Phimme}), expand $\Phi_{2,2}(x,i)$ in $x$ to order $x^4$ for arbitrary $i$,
and obtain the resonant torque to order $x^6$ as
\begin{eqnarray}
\label{t22a}
T_{2,2}(i) \simeq & &- \frac{3 \pi^2}{8} \mu^2 \Sigma a^4 \Omega_{\rm b}^2 x_{\rm res}^4 \cos^8{(i/2)} \times \\[5pt]
                 & &\left[9 + 5 x_{\rm res}^2(9 -14 \cos{(i)} + 7 \cos{(2 i)}) \right]. \nonumber
\end{eqnarray}
where $x_{\rm res}$ is the value $x$ at the resonance location $x_{\rm res}=2^{-2/3} (1-\mu)^{1/3}$. To lowest order in $x_{\rm res}$, the torque $T_{2,2}(i) \propto \cos^8{(i/2)}$. This lowest order term is generally not a good approximation to the actual value of $T_{2,2}$ because $x_{\rm res}$ is not very small. Consequently we include the next order term.

To more accurately evaluate the torque in Equation (\ref{tgt}), we also apply  a numerical approach. In this approach, we determine $\Phi_{2,2}(x,i)$ by carrying out the 2D integral in Equation (\ref{Phimme}) by numerical means with $m=2$.

\begin{figure*}
  \includegraphics[width=0.47\textwidth]{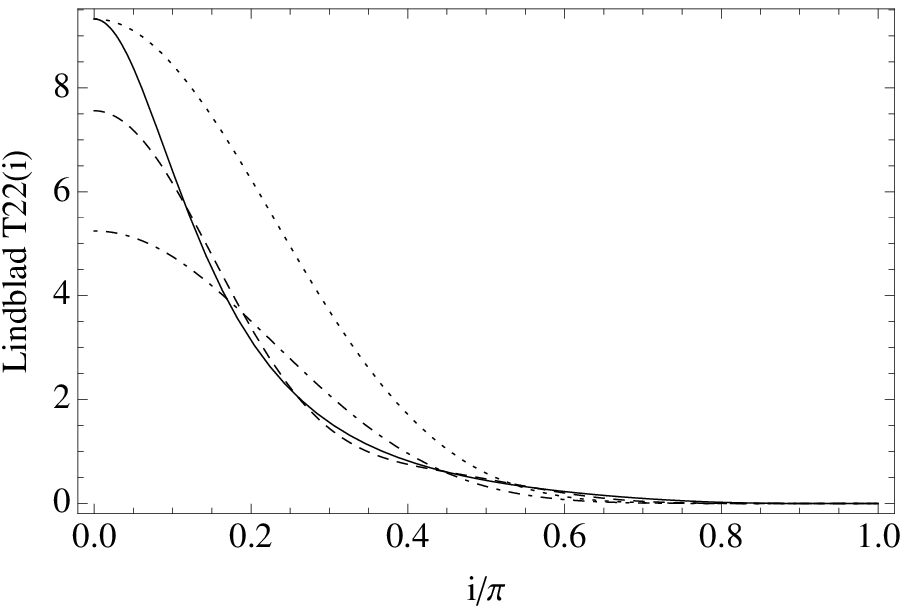}
  \hspace{0.06\textwidth}
  \includegraphics[width=0.47\textwidth]{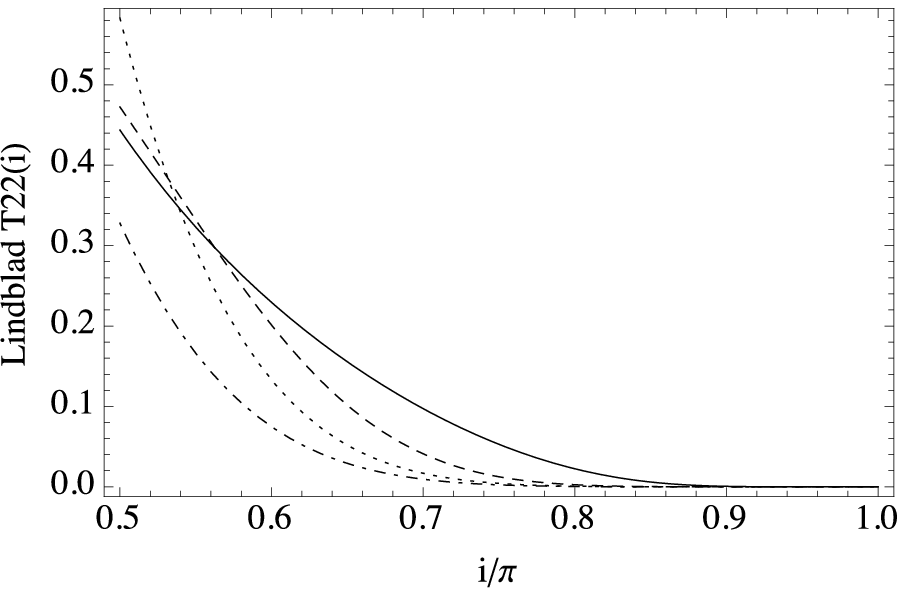}
  \caption{Resonant torque as a function of binary-disk inclination due to the 2:1 inner Lindblad resonance at  radius of $r=a/2^{2/3}$ for a small mass perturber of mass fraction $\mu \ll 1$.  The right panel shows a close-up view of the left panel at larger inclination angles.} The torque is normalized by  $-\Sigma(r_{\rm res})  a^4 \Omega_{\rm b}^2 \mu^2$. The solid line plots the torque based on the  potential component $\Phi_{2,2}$ that is accurately evaluated using a numerical integration in Equation (\ref{Phimme}) with $m=2$, the dashed line is the analytic series approximation given by Equation (\ref{t22a}),  the dotted-dashed line is the analytic series approximation given by Equation (\ref{t22a}) taken to lowest order (order $x_{\rm res}^4$), and the dotted line is  $\cos^8{(i/2)}$ times the torque at $i=0$ for the solid line.
  \vspace{0.1in}
  \label{Tq0}
\end{figure*}

\begin{figure*}
  \includegraphics[width=0.47\textwidth]{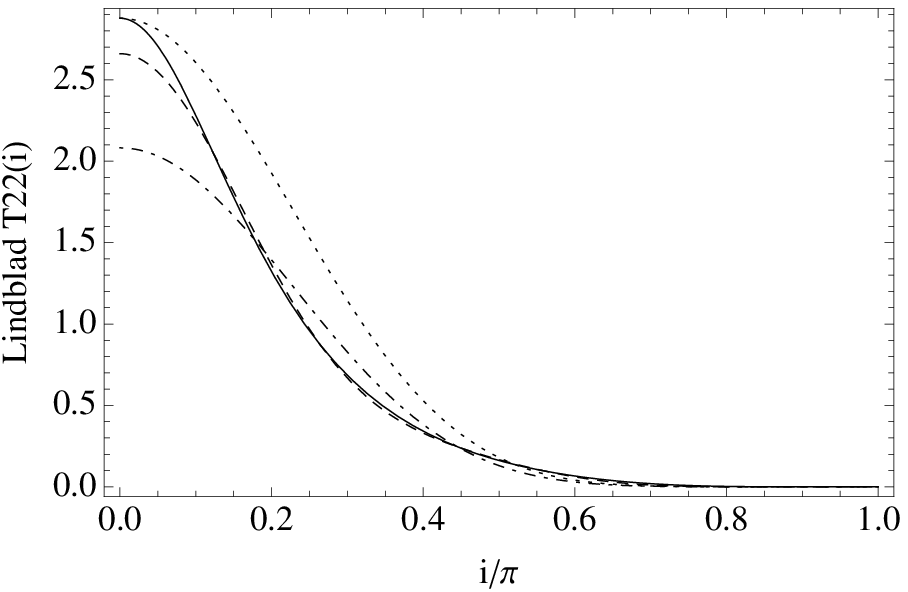}
  \hspace{0.06\textwidth}
  \includegraphics[width=0.47\textwidth]{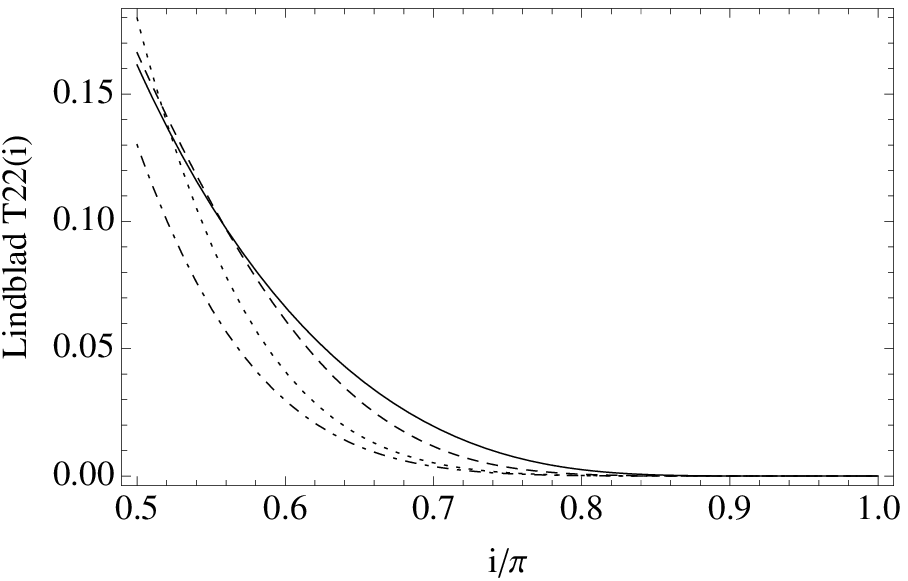}
  \caption{Same as Figure \ref{Tq0}, but for  an equal mass binary $\mu=0.5$. 
    \vspace{0.1in}}
  \label{Tq1}
\end{figure*}

Figures \ref{Tq0} and  \ref{Tq1} plot as solid lines the variations of the 2:1 resonant torque as a function of inclination angle $i$ for small binary mass ratio and unit binary mass ratio, respectively. In each figure, the right panel is a more detailed view of the left panel at higher inclination angles.
In both cases, for $0< i < \pi/2$ the torque drops rapidly with inclination angle, faster than $\cos^8{(i/2)}$. Note that in these figures, the dotted lines that decrease as $\cos^8{(i/2)}$ are normalized by the accurate numerically determined value of $T_{2,2}$ at $i=0$ (solid lines). The lowest order (order  $x_{\rm res}^4$) torque in Equation (\ref{t22a}) ,
which is $\propto \cos^8{(i/2)}$, is plotted as 
dotted-dashed lines. The agreement of the dotted-dashed lines with the solid lines is poorer at small and large inclination angles. The figure shows the increased accuracy obtained including the next order correction (dashed lines) in Equation  (\ref{t22a}).

At larger inclination angles, the torque declines more slowly  than  $\cos^8{(i/2)}$. The analytic approximation given by Equation (\ref{t22a}) is fairly accurate  for $i \ga \pi/6$. The torque reaches half the coplanar value at about $i \simeq 0.45$ ($26^{\circ}$) and $i \simeq 0.6$ ($34^{\circ}$), in the small and unit mass ratio cases, respectively. For $i=\pi/2$, the torques decline by about a factor of 20 in both cases.  For $i=\pi$, the counter-rotating case, the torque is zero in both cases, as expected (see Section \ref{intro}). 

Close to counteralignment where $i \simeq \pi$, Equation (\ref{t22a}) predicts that $T_{2,2} \propto (i-\pi)^8$. Although Equation (\ref{t22a}) gives the proper lowest order dependence on $i$, the coefficient of  $(i-\pi)^8$ converges slowly in $x_{\rm res}$ near counteralignment. To obtain an accurate analytic expression for $T_{2,2}$ in this limit, we expand  Equation (\ref{Phimme})  and its $x$-derivative in $i$ about $i=\pi$ for $x=x_{\rm res}$ and apply the results to the torque Equation (\ref{tgt}). We then obtain an accurate expression in the nearly counteraligned case,
\begin{equation}
T_{2,2}(i) \simeq - \beta \, \mu^2 \, \Sigma \, a^4 \, \Omega_{\rm b}^2 \, (i-\pi)^8,
\label{T22i}
\end{equation}
where $\beta = 17.9$ for a very small mass perturber, $\mu \ll1$, and $\beta = 0.49$ for an equal mass binary, $\mu=0.5$. We have tested the accuracy of this approximation by comparing it with the numerical evaluation of the torque based on Equations (\ref{Phimme})  and  (\ref{tgt}) with $m=2$. The fractional errors of Equation (\ref{T22i}) are $\sim$15\% for $i \simeq 0.97 \pi$ and go to zero as $i$ approaches $\pi$.

\subsection{\texorpdfstring{$m>2$}{m=2} Lindblad resonances}
In the coplanar case $i=0$, Lindblad torques due to $\ell=m > 2$ provide even stronger torques per unit disk surface density than the case for $\ell=m = 2$. Their greater strength is due to the fact they lie closer to the perturber, at radial distance $\sim a/m$ from the orbit of the perturber. However, they may be excluded from the disk by the tidal truncation effects of the $\ell=m = 2$  inner Lindblad resonance, as is commonly the case for prograde coplanar binary-disk systems with mass ratios of order unity, resulting in the $m=2$ resonance being the strongest to affect the disk. In the prograde coplanar case for a small mass perturber $\mu \ll 1$, the torques  generally increase quadratically with $m$ for $m \la a/H$ for disk thickness $H$. For larger values of $m$ and small perturber mass $\mu \ll 1$, beyond the so-called torque cutoff, the torque decreases due the finite thickness and pressure effects of the disk (\citealt{Goldreich80}, \citealt{Artymowicz93}).

For the case that the disk and binary orbit are misaligned, the high $m$ resonances are not located as close  to the perturber, when averaged over an orbit. Consequently, there may be changes in the behavior of the torques $T_{m,m}$ for large $m$. In addition, for the lowest order term in $x$, the torque $T_{m,m} \propto \cos^{4 m}{(i/2)}$. So although these higher $m$ resonances are stronger for $i=0$ than the $m=2$ resonance, their torques decline more rapidly with inclination. We ignore disk pressure or finite thickness effects. We calculate the torques for the  $m>2$ resonances by numerically integrating Equation (\ref{Phimme}) and applying Equation  (\ref{tgt}).

\begin{figure}
  \includegraphics[width=\columnwidth]{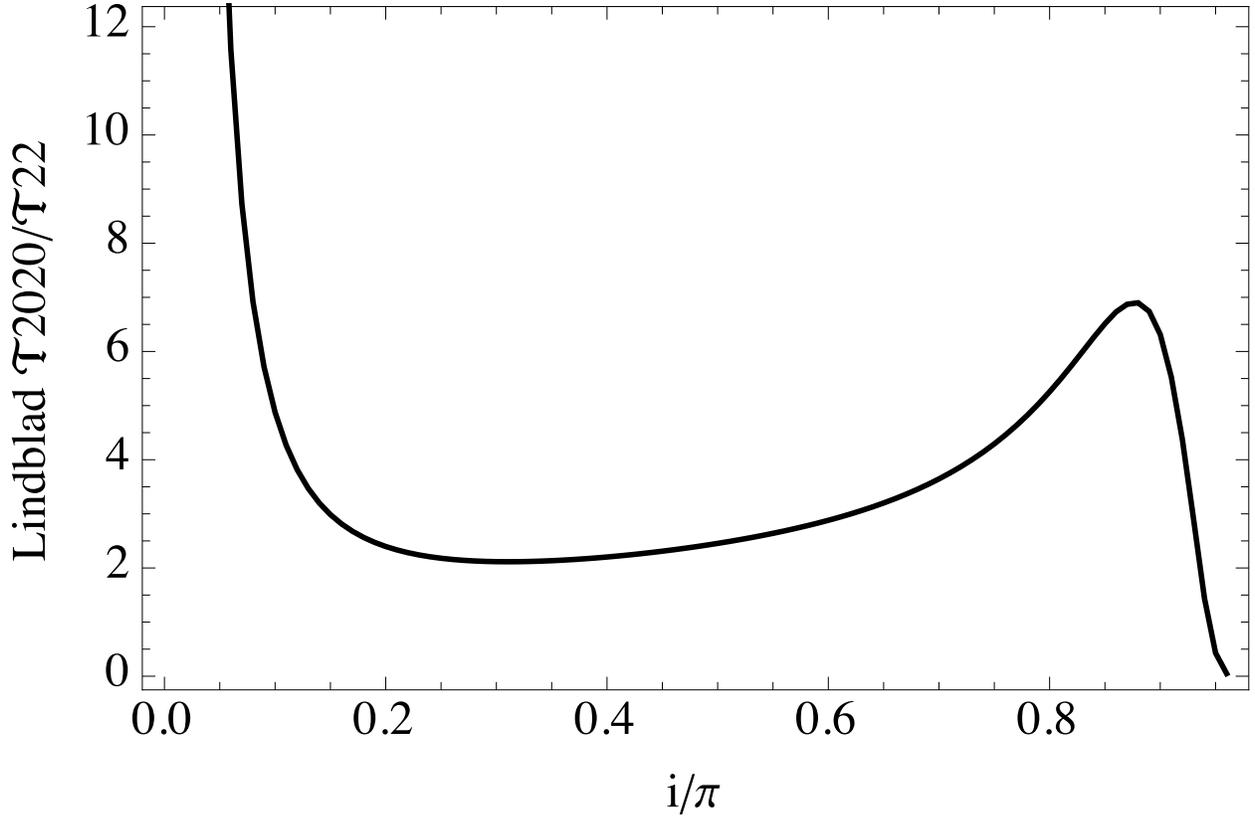}
  \hspace{0.06\textwidth}
      \caption{Torque ratio ${\cal T}_{20,20}/{\cal T}_{2,2}$ for $\mu \ll 1$ is plotted as a function of inclination angle in radians.
    \vspace{0.1in}}
  \label{TmT2q0i}
\end{figure}

We define the torque per unity disk surface density as ${\cal T}_{m, m}= T_{m, m}/\Sigma(r_{\rm res})$. 
   Figure \ref{TmT2q0i} plots ratio ${\cal T}_{20,20}/{\cal T}_{2,2}$ as a function of inclination $i$ 
   with a small mass perturber, $\mu \ll 1$.  This ratio is the same as the ratio of the torques for a constant surface density disk.
   The plot shows that for $0.2 \pi \la i \la 0.7 \pi$, the $m=2$ resonant torque  is only factor of a few smaller than the $m=20$ torque,  that is close
   in value to the maximum resonant torque across all $m$. This result suggests that a viscous disk may be able to overrun all resonances if the viscous torque exceeds $T_{2,2}$ by a factor of a few for a small mass perturber in this inclination range. For $ 0.7 \pi  \la i \la 0.9 \pi$, the ratio increases somewhat to about 7. For larger inclinations, the ratio drops and the $m=2$ torque is the strongest in the nearly counter-rotating case for $i \ga 0.95 \pi$.

\begin{figure*}
  \includegraphics[width=0.47\textwidth]{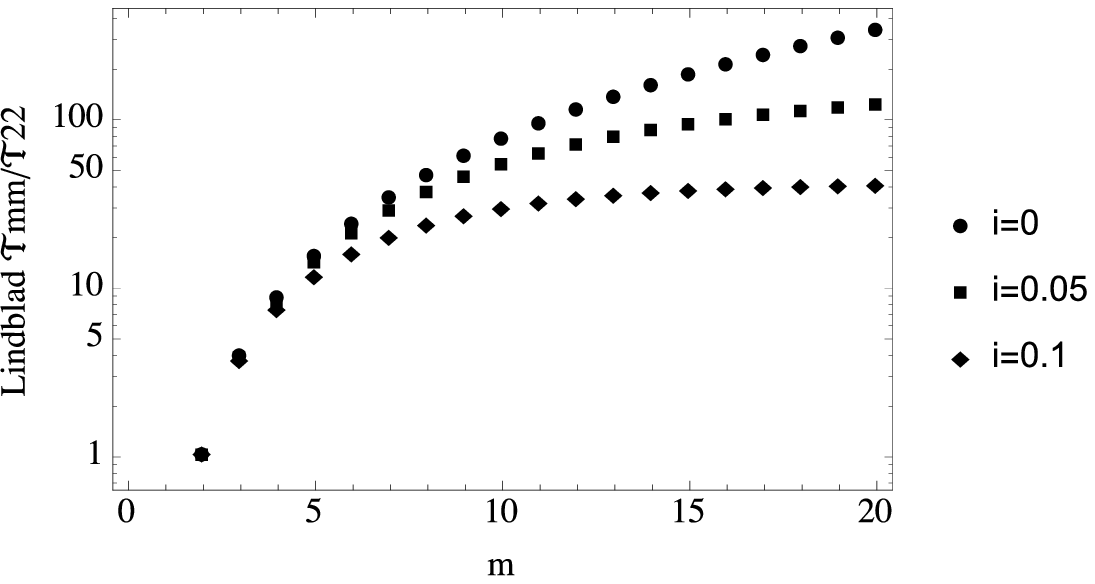}
  \hspace{0.06\textwidth}
  \includegraphics[width=0.47\textwidth]{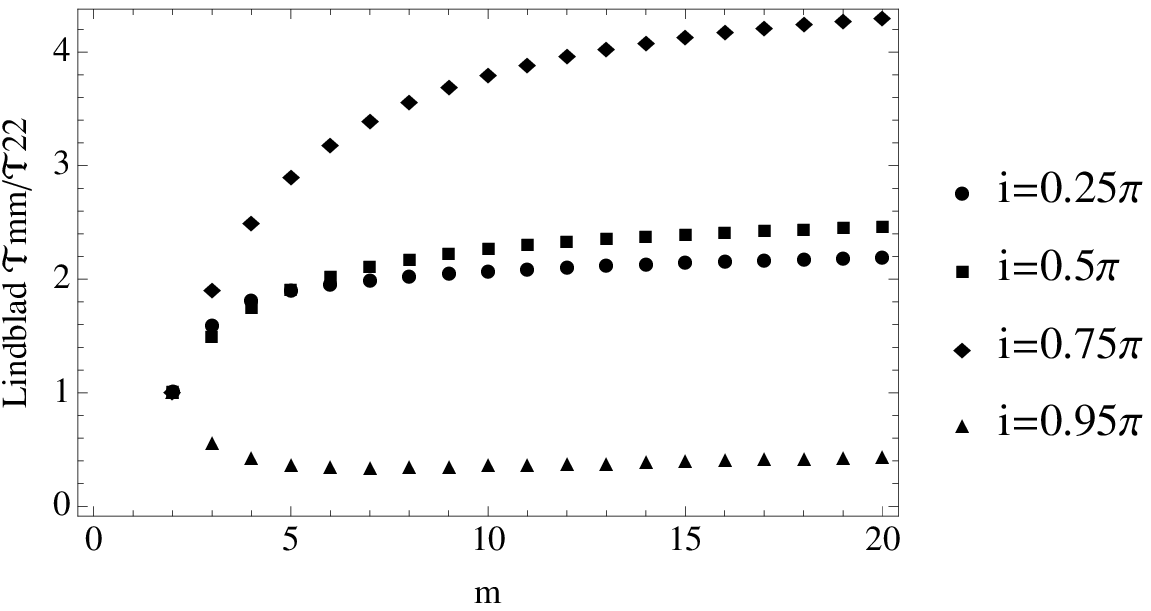}
     \caption{Torque ratio ${\cal T}_{m,m}/{\cal T}_{2,2}$ for $\mu \ll 1$ is plotted as a function of azimuthal wavenumber $m$ for different binary--disk inclination angles $i$.
    \vspace{0.1in}}
  \label{Tmmu0}
\end{figure*}

Figure \ref{Tmmu0} plots the ratio  ${\cal T}_{m,m}/{\cal T}_{2,2}$  as a function of $m$ for different small inclinations with a small mass perturber, $\mu \ll 1$. 
In the left panel, for inclination $i \sim 0.1$, ${\cal T}_{m,m}/{\cal T}_{2,2}$ 
 reaches roughly half its maximum value at $m \sim 0.8/i$ (with $i$ in radians).
 For a fixed value of large $m$, the variation of ${\cal T}_{m,m}/{\cal T}_{2,2}$  with inclination is non-monotonic. ${\cal T}_{m,m}/{\cal T}_{2,2}$  initially decreases strongly with $i$ for small $i$ (left panel), then grows somewhat before dropping to zero for the counter-rotating case (right panel), as is consistent
with Figure \ref{TmT2q0i}.
 The higher $m$ torques are limited in strength for misaligned binary--disk systems. Notice that for angles, $i > 0.05$, 
 ${\cal T}_{m,m}/{\cal T}_{2,2}$ is nearly constant with $m$ near $m=20$ and is nearly maximum for $m=20$. (For the case of $i=0.75 \pi$, ${\cal T}_{20,20}$ is within 3\% of its peak value that occurs for $m>20$.) For the nearly counter-rotating case of $i= 0.95 \pi$, the  $m=2$ torque is the largest torque contributor.

\begin{figure*}
  \includegraphics[width=0.47\textwidth]{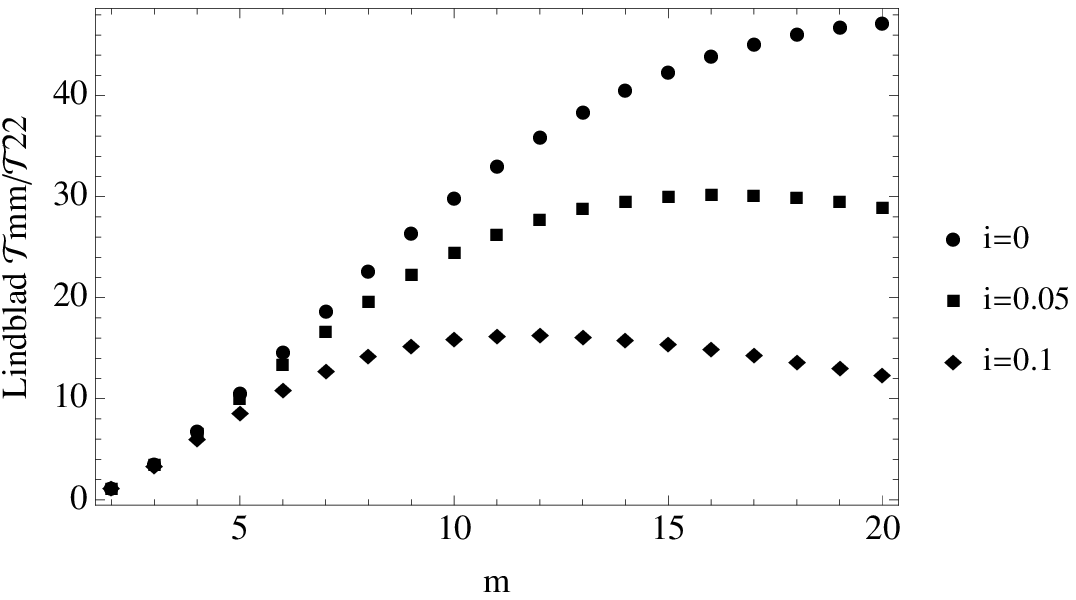}
  \hspace{0.06\textwidth}
  \includegraphics[width=0.47\textwidth]{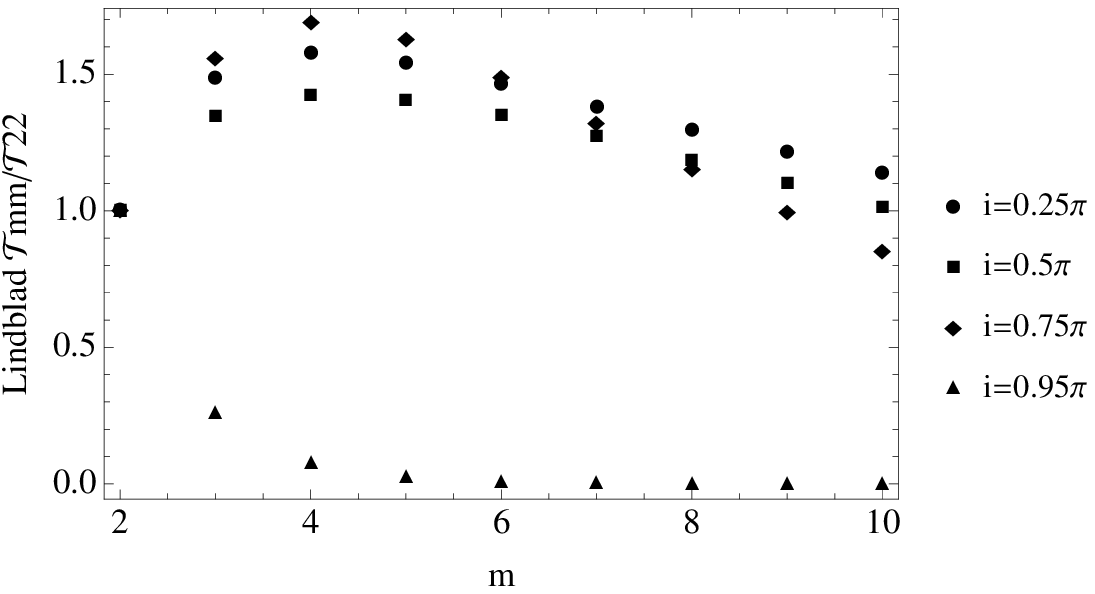}
     \caption{Torque ratio ${\cal T}_{m,m}/{\cal T}_{2,2}$ for $\mu = 0.1$ is plotted as a function of azimuthal wavenumber $m$ for different binary--disk inclination angles $i$.
    \vspace{0.1in}}
  \label{Tmmu0p1}
\end{figure*}

The behavior of the torque variation for intermediate mass perturbers
 is plotted in Figure \ref{Tmmu0p1}. The left panel of Figure \ref{Tmmu0p1} 
shows that for fixed $m$ and small inclinations, the torques relative to the $m=2$ torque
decrease with inclination, as in Figure  \ref{Tmmu0}. However, this torque  variation with inclination is more complicated at higher inclinations. 
The right panel of Figure \ref{Tmmu0p1} shows that for $m=4$, the torques vary non-monotonically respect to the $m=2$ torque
 as the inclination increases.  On the other hand, for  $m=20$ the torque ratio decreases monotonically with inclination.
For intermediate inclinations $0.25 \pi < i < 0.75 \pi$, the Lindblad torque ratio for $m>2$ does not increase as much as the $\mu \ll 1$ case. Typically, the $m>2$ torques are at most $\sim$60\% higher than $T_{2,2}$ for $i > 0.25 \pi$. The maximum torque occurs for $m = 4$.  
In the nearly counter-rotating case $i=0.95 \pi$ the torque declines rapidly with $m$ for $m>2$.

\begin{figure*}
  \includegraphics[width=0.47\textwidth]{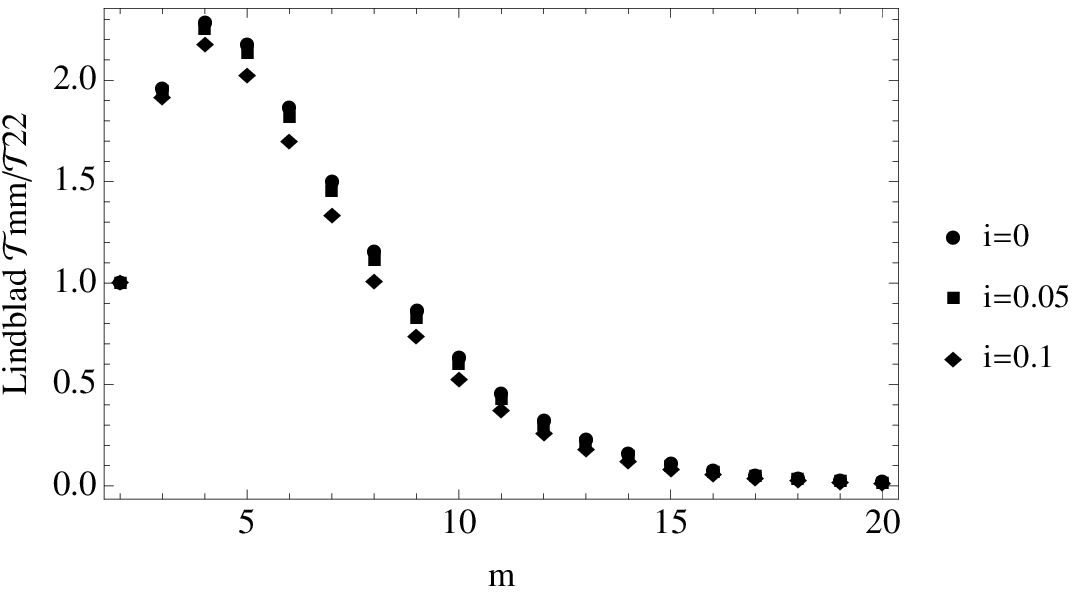}
  \hspace{0.06\textwidth}
  \includegraphics[width=0.47\textwidth]{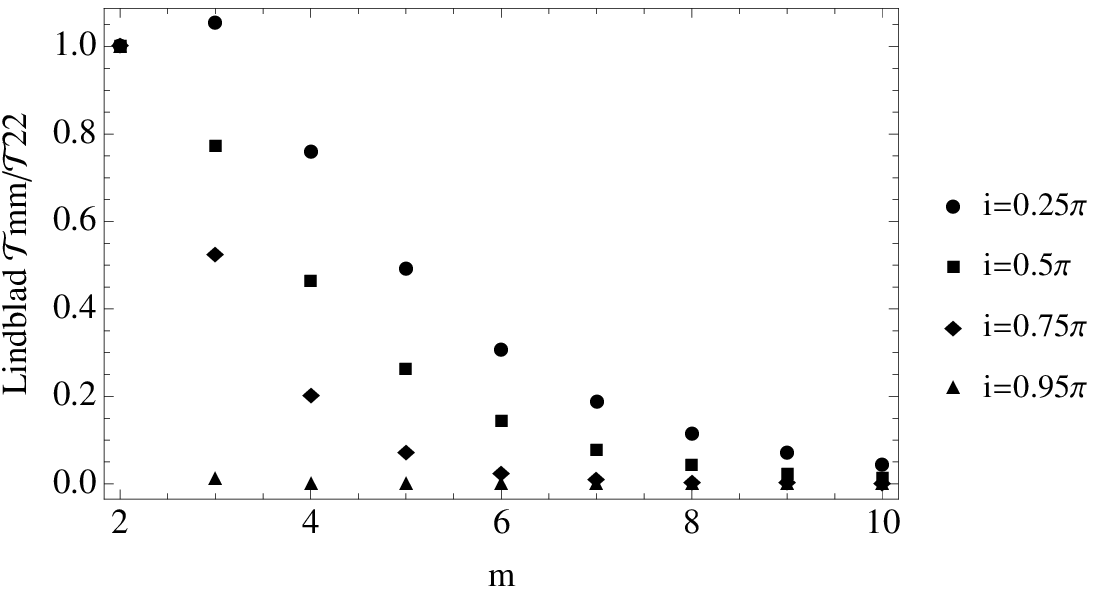}
     \caption{Torque ratio  ${\cal T}_{m,m}/{\cal T}_{2,2}$ for an equal mass binary $\mu = 0.5$ is plotted as a function of azimuthal wavenumber $m$ for different binary--disk inclination angles $i$.
    \vspace{0.1in}}
  \label{Tmmu0p5}
\end{figure*}

As seen in Figure \ref{Tmmu0p5} for an equal mass binary and inclinations above a certain value, $0.25 \pi < i $,  ${\cal T}_{2,2}$ dominates or is nearly the same as ${\cal T}_{m,m}$ for  $m>2$. From these results, we conclude that the higher $m$ resonances become much weaker with increasing inclination. Their torques decrease in strength relative to the $m=2$ case for higher secondary mass $M_2 < M_1$ perturbers.

\section{Comparing the tidal and viscous torques}
\label{comp}
For tidal truncation to occur, the resonant torque must be stronger than the viscous torque. In this section we compare the tidal truncation torque to the disk viscous torque. We assume the disk is planar (unwarped), but misaligned to the binary orbit. The internal viscous torque \citep[e.g.][]{Pringle81} is given by
\begin{equation}
T_{\nu} = 3{\rm \pi}\alpha \left(\frac{H}{r} \right)^2 \Sigma \Omega^2 r^4,
\label{gnu}
\end{equation}
where  $\alpha$ the dimensionless viscosity parameter and $H$ is the disk thickness. Tidal truncation occurs at or interior to the innermost resonance (closest to the central object) whose resonant torque dominates over the viscous torque.

To illustrate the possible outcomes of tidal truncation, we describe below a few representative cases. In these cases the resonant torque was determined by  carrying out the  2D integral in Equation (\ref{Phimme}) for $\Phi_{m,m}$ and $\partial_x \Phi_{m,m}$  by numerical means and then applying them to Equation (\ref{tgt}).

Consider a binary with perturber mass fraction $\mu=0.1$,  $H/r=0.05$, and $\alpha=0.1$. Applying the truncation condition that the Lindblad torque $T_{2,2}$ be greater than $T_{\nu}$, we find that the disk can expand beyond the 2:1 resonance if $i  \ga 2.02$ ($116^{\circ}$). As we see in Figure \ref{Tmmu0p1}, overrunning all Lindblad resonances requires overcoming a $\sim 60\%$ stronger torque due to the $m=4$ torque. This can occur for a more inclined disk with $i \ga 2.18$ ($125^\circ$). The small increase in angle to overrun all resonances is due to the rapid decline in torque with inclination. 
 
For  a  cooler disk, overcoming the tidal barrier requires that the disk be closer to counter-rotating. For a binary with perturber mass fraction $\mu=0.1$,  $H/r=0.01$, and $\alpha=0.1$, the disk can expand beyond the 2:1 Lindblad resonance if $i  \ga 2.65$ ($152^{\circ}$).

For an equal mass binary (perturber mass fraction $\mu=0.5$),  $H/r=0.1$, and $\alpha=0.1$, the disk can expand beyond the 2:1 resonance if $i  \ga 2.34$ ($134^{\circ}$). In this case, the tidal torque is so powerful that the inclination must be large, although not fully counter-aligned,  for the viscous torques to dominate. Figure \ref{Tmmu0p5} suggests that the 2:1 ($m=2$) resonance is the strongest resonance at this misalignment angle. The other resonances would then not be able to truncate such a disk. 
 
\section{Discussion}
We have made some approximations in this analysis. We have applied the linear theory of resonant torques. The complication that arises here is that orbit crossings can occur, as is well known in the aligned $i=0$ case, that may prevent resonances such as the 2:1 resonance from residing within the disk. At large inclinations, the resonant torques become weaker and more linear and the model we have described is more applicable.

In the counteraligned limit, the effects of the perturber on the disk orbits are considerably weaker, but must be important at the Bondi radius of the perturber. At such radii, the assumption we made that Keplerian orbits occur about mass $M_{\rm c}$ is violated. This condition suggests that with small $\mu$, this model is inapplicable near $i \simeq \pi$ for the values of $m \ga 1/\mu$.

Another limitation of the current study is that we have assumed that the disks are circular (apart from tidal distortions). Eccentric disks may be able to more easily extend to larger radii than is indicated in the present study, due to excursions of gas to the disk apocenter. In \cite{Martin14a}, we found in simulations that gas capture by a binary companion appears to occur more easily if the disk is eccentric. Furthermore, substantially inclined disks can be subject to Kozai-Lidov oscillations that generate disk eccentricity  \citep{Martin14b}. Consequently, the effects of disk inclination may naturally bring about additional effects due to disk eccentricity that in turn allows for greater disk extension.

We have assumed the binary orbit is circular. Eccentric orbits of binary systems give rise to additional resonances that can truncate a disk \citep{Artymowicz94}. We have also ignored the effects of vertical forcing of the disk by the binary. 

In some systems (e.g. Roche-lobe overflow from a donor star), the disk is fed by gas that lies in the binary plane. In such systems, it is unlikely that the disk truncation is strongly affected by inclination effects. However, there are tilting instabilities that can misalign the disk from the binary plane (e.g., radiation warping, \citealt{Pringle96}). 

In other systems, the disk misalignment to the binary plane occurs at disk formation (e.g., young wide binaries such as HK Tau \citealt{Jensen14}, Be/X-ray binaries \citealt{Martin14a}; or SMBH binary disks, \citealt{Nixon13}). 
If the disk is formed
smaller than its tidal limit, then viscous evolution could cause the disk to expand to its tidal limit.
In these systems it is important to consider whether the tilt decays quickly. An inclined disk in a binary system is subject to inclination decay due to viscous dissipational effects (\citealt{Papaloizou95}; \citealt{Larwood96}; \citealt{Lubow00}). For large $\alpha$ (say $\gtrsim 0.3$) the ratio of the vertical and azimuthal viscosities in a warped disk  is of order unity \citep{Papaloizou83} and therefore the disk is able to spread significantly before any alignment takes place. For $\alpha \lesssim 0.1$ the disk may instead align more quickly at the outer edge, requiring some instability (as mentioned above) to maintain the tilt.

If there is a strong reduction in the tidal torques with inclination, gas in the disk may be captured by the companion object, possibly flow into a circumbinary disk, or even leave the binary. As noted above, this process may be facilitated by disk elongation if eccentricity is present, in addition to the weakening of the tidal torques with inclination that we have investigated in this paper. In a SMBH binary that is coplanar with a circumbinary disk, the smaller mass black hole accretes most of the available gas, since its orbit lies closer to the circumbinary disk which is feeding the system \citep{Artymowicz96}. If the circumsecondary disk is significantly misaligned to the binary orbit, the reduced tides might allow mass transfer  from the secondary to the primary black hole.

\section{Conclusions}
We have analyzed the effects of binary-disk inclination on Lindblad resonant tidal torques that act on a disk that surrounds a member of a binary system. We considered the binary orbit and the disk to be circular. We focused on the 2:1 inner Lindblad resonance with azimuthal wavenumber $m=2$ that dominates the tidal truncation of coplanar disks by a binary that rotates in the same sense (prograde). For that resonance, the torque declines rapidly with inclination angle (see Figures \ref{Tq0} and \ref{Tq1}), faster (slower) than $\cos^8{(i/2)}$ for inclination angle $0<i<\pi/2$ ($\pi/2<i<\pi$). Close to counter-rotation, $i=\pi$, the torque declines rapidly to zero as $(i-\pi)^8$.

An inclined disk is  expected to be somewhat larger than the corresponding coplanar $i=0$ disk due to the reduction in the strength of the Lindblad tidal torques. Viscous torques can dominate the 2:1 Lindblad resonance if the binary is counter-rotating ($i=\pi$). But they can also dominate for smaller tilts, provided the disk is sufficiently viscous, even for situations where the  Lindblad torque would strongly dominate for the aligned case ($i=0$). Higher order $m>2$ resonances may also play a role in truncating the disk. However, we have found that such resonances are weakened considerably with inclination relative to the $m=2$ resonance (Figures \ref{TmT2q0i}  - \ref{Tmmu0p5}). They also weaken  relative to the $m=2$ resonance with  increasing perturber mass fraction in going from $\mu \ll 1$ to the equal mass binary case of $\mu=0.5$. This results suggest that under some circumstances (sufficiently high viscosity and sufficiently small perturber mass), a disk that would be strongly tidally truncated for $i=0$ can overrun all Lindblad resonances, even if is not fully counter-rotating ($i< \pi$). 

\acknowledgments{SHL acknowledges support from NASA grant NNX11AK61G. CJN acknowledges support provided by NASA through the Einstein Fellowship Program, grant PF2-130098. RGM's support was provided under contract with the California Institute of Technology (Caltech) funded by NASA through the Sagan Fellowship Program. We thank the referee for suggesting improvements in the description of the results.}

\bibliographystyle{apj} 
\bibliography{ref}

\end{document}